\def\titlep{Branching laws for polynomial endomorphisms 
in CAR algebra for fermions,
uniformly hyperfinite algebras and Cuntz algebras}
\documentclass[11pt]{article}
\usepackage{graphicx,ifthen}
\usepackage{amsmath}

\font\germ=eufm10 at12pt

\def\goth#1{\hbox{\germ#1}}

\newcommand{\qed}{\hbox{\rule[-2pt]{3pt}{6pt}}}
\newcommand{\qedh}{\hfill\qed \\}

\newcommand{\vv}{\vspace{.3in}}

\newcommand{\vep}{\varepsilon}

\newtheorem{Thm}{Theorem}[section]

\newtheorem{defi}[Thm]{Definition}
\newtheorem{lem}[Thm]{Lemma}

\newtheorem{prop}[Thm]{Proposition}

\def\toprule{\hline}
\def\botrule{\hline}
\def\colrule{\hline}
\newcommand{\ww}{\vv\noindent}

\def\cal#1{\mathcal #1}
\def\con{{\cal O}_{N}}
\def\edot{=1,\ldots,N}
\def\pr{\noindent{\bf Proof.}\,\,}

\def\scm#1{S({\bf C}^{N})^{\otimes #1}}

\def\evp{eventually periodic}
\def\nset#1{\{1,\ldots,N\}^{#1}}

\def\co#1{{\cal O}_{#1}}
\def\brl{branching law}
\def\bfsnl{{\rm BFS}_{N}(\Lambda)}

\def\enda{{\rm End}{\cal A}}

\addtocounter{footnote}{1}
\def\sftt#1{
\setcounter{equation}{0}
\addtocounter{footnote}{1}
\section{#1}
}

\def\ssft#1{\subsection{#1}}

\begin{document}

%
%
\def\autherp{Mitsuo Abe and Katsunori Kawamura}
\def\autherr{Mitsuo Abe}
\def\autherq{Katsunori Kawamura}
\def\emailp{abe@kurims.kyoto-u.ac.jp}
\def\emailq{kawamura@kurims.kyoto-u.ac.jp}
\def\addressp{Research Institute for Mathematical Sciences \\
Kyoto University, Kyoto 606-8502, Japan}
\def\addressq{College of Science and Engineering Ritsumeikan University,\\
1-1-1 Noji Higashi, Kusatsu, Shiga 525-8577, Japan
}

\def\infw{\Lambda^{\frac{\infty}{2}}V}
\def\zhalfs{{\bf Z}+\frac{1}{2}}
\def\ems{\emptyset}
\def\vac{|{\rm vac}\rangle _{+}}
\def\dvac{|{\rm vac}\rangle _{-}}
\def\ovac{|0\rangle}
\def\tovac{|\tilde{0}\rangle}
\def\expt#1{\langle #1\rangle}
\def\zph{{\bf Z}_{+/2}}
\def\zmh{{\bf Z}_{-/2}}
\def\brl{branching law}
\def\bfsnl{{\rm BFS}_{N}(\Lambda)}
\def\scm#1{S({\bf C}^{N})^{\otimes #1}}
\def\mqb{\{(M_{i},q_{i},B_{i})\}_{i=1}^{N}}
\def\zhalf{\mbox{${\bf Z}+\frac{1}{2}$}}
\def\zmha{\mbox{${\bf Z}_{\leq 0}-\frac{1}{2}$}}
\newcommand{\mline}{\noindent
\thicklines
\setlength{\unitlength}{.1mm}
\begin{picture}(1000,5)
\put(0,0){\line(1,0){1250}}
\end{picture}
\par
 }

%

%
%
\begin{center}
{\Large \titlep}

\ww
\autherr
\footnote{\emailp}

\quad
\\

\noindent
{\it \addressp}

\quad
\\

\autherq
\footnote{\emailq}

\quad
\\

\noindent
{\it \addressq}
\quad \\
\quad \\
\quad \\
\end{center}

%
%
\begin{abstract}
Previously,  we have shown that the CAR algebra for fermions 
is embedded in the Cuntz algebra ${\cal O}_{2}$ in such a way 
that the generators are expressed in terms of 
polynomials in the canonical generators of the latter, and 
it coincides with the $U(1)$-fixed point subalgebra 
${\cal A}\equiv {\cal O}_{2}^{U(1)}$ of ${\cal O}_{2}$ for
the canonical gauge action.
Based on this embedding formula, 
some properties of ${\cal A}$ are studied in detail by restricting 
those of ${\cal O}_{2}$. 
Various endomorphisms of ${\cal O}_{2}$, 
which are defined by polynomials in the canonical generators, 
are explicitly constructed, and transcribed into those of ${\cal A}$.
Especially, we investigate
branching laws for a certain family of such endomorphisms with respect to 
four important representations, i.e.,
the Fock representation, the infinite wedge representation and their duals.
These endomorphisms are completely classified by their branching laws. 
As an application, we show that  the reinterpretation 
of the Fock vacuum as the Dirac vacuum
is described in representation theory through a mixture of fermions.
\end{abstract}



%
%
\sftt{Introduction}
\label{section:first}
In previous papers \cite{AK1,AK06,AK05,AK2}, 
we have presented a recursive construction of the CAR 
(canonical anticommutation relation) algebra for fermions in terms of 
the Cuntz algebra $\co{2}$ and shown that 
it may provide us a useful tool to study properties of fermion systems 
by using explicit expressions in terms of generators of the algebra. 

Let ${\cal A}_{0}$ be the algebra generated by 
$a_{n},a_{n}^{*}$ for $n\in{\bf N}\equiv \{1,2,3,\ldots\}$
which satisfy the canonical anticommutation relations (=CAR):
%
%
\begin{equation}
\label{eqn:fereq}
a_{n}a_{m}^{*}+a_{m}^{*}a_{n}=\delta_{n,m}I,\quad
a_{n}^{*}a_{m}^{*}+a_{m}^{*}a_{n}^{*}=a_{n}a_{m}+a_{m}a_{n}=0
\quad(n,m\in {\bf N}).
\end{equation}
The $*$-algebra ${\cal A}_{0}$ always has unique C$^{*}$-norm $\|\cdot\|$ 
and the completion ${\cal A}$ of ${\cal A}_{0}$ with respect to
$\|\cdot\|$ is called the {\it CAR algebra} in theory of operator algebras.
Let $s_{1},s_{2}$ be canonical generators of $\co{2}$.
Then
%
%
\begin{equation}
\label{eqn:recursive}
a_{1}\equiv s_{1}s_{2}^{*},\quad
a_{n}\equiv \sum_{J\in\{1,2\}^{n-1}}
s_{J}s_{1}s_{2}^{*}\beta_{2}(s_{J})^{*} \quad(n\geq 2)
\end{equation}
satisfy (\ref{eqn:fereq})
where $s_{J}\equiv s_{j_{1}}\cdots s_{j_{k}}$ for $J=(j_{1},\ldots,j_{k})$
and $\beta_{2}$ is the automorphism of $\co{2}$ defined by 
$\beta_{2}(s_{i})\equiv (-1)^{i-1}s_{i}$ for $i=1,2$ \cite{AK1}.
Furthermore, $C^{*}\langle\{a_{n}\in\co{2}:n\in {\bf N}\}\rangle$
coincides with a fixed-point subalgebra 
$\co{2}^{U(1)}\equiv \{x\in \co{2}: \mbox{for all }z\in U(1),\,
\gamma_{z}(x)=x\}$ of $\co{2}$ for
the canonical gauge action $\gamma$
defined by $\gamma_{z}(s_{i})=zs_{i}\ (i=1,\,2)$ with $z\in U(1)$.
Define the linear map $\zeta$ on $\co{2}$ by
%
%
\begin{equation}
\label{eqn:rec}
\zeta(x)\equiv s_{1}xs_{1}^{*}-s_{2}xs_{2}^{*}\quad(x\in \co{2}).
\end{equation}
Then we have $a_{n}=\zeta(a_{n-1})$ for each $n\geq 2$.
In this sense, $\{a_{n}\}_{n\in {\bf N}}$ in (\ref{eqn:recursive})
is called the {\it recursive fermion system (=RFS)} in $\co{2}$.
Remark that (\ref{eqn:recursive}) is a finite sum  giving a 
purely algebraic embedding, 
while it is also a (continuous) embedding of the C$^{*}$-algebra
${\cal A}$ into $\co{2}$.
In contrast with the so-called boson-fermion correspondence \cite{MJD, Oko01},
in which the map between bosons and fermions
depends on their representations, the generators of 
fermions are always written by
noncommutative homogeneous polynomials 
($\in{\bf C}[s_{1},s_{2},s_{1}^{*},s_{2}^{*}]$) in the
canonical generators of $\co{2}$ without use of any representation.

We have also shown \cite{AK05} that it is possible to generalize 
this recursive construction to the algebra for the FP ghost fermions 
in string theory by introducing a $*$-algebra called 
the pseudo-Cuntz algebra suitable for actions on 
an indefinite-metric state vector space. 
We have found that, according to embeddings of the FP ghost algebra 
into the pseudo-Cuntz algebra with a special attention to the zero-mode 
operators, unitarily inequivalent representations 
for the FP ghost are obtained from 
a single representation of the pseudo-Cuntz algebra. 

Based on the embedding formula (\ref{eqn:recursive}), 
some properties of ${\cal A}$ are studied in detail by restricting 
those of ${\cal O}_{2}$. 
For this purpose, we start with branching laws of representations of $\con$
restricted on $\con^{U(1)}$.
%
%
\begin{defi}[Permutative representation of $\con$]
\label{defi:first}
Let $s_{1},\ldots,s_{N}$ be canonical generators of $\con$.
\begin{enumerate}
\item
For $J=(j_{l})_{l=1}^{k}\in\nset{k}$,
$P(J)$ is the class of representations $({\cal H}, \pi)$ of $\con$ with 
a cyclic unit vector $\Omega\in {\cal H}$
such that $\pi(s_{J})\Omega=\Omega$
and $\{\pi(s_{j_{l}}\cdots s_{j_{k}})\Omega\}_{l=1}^{k}$
is an orthonormal family in ${\cal H}$
where $s_{J}\equiv s_{j_{1}}\cdots s_{j_{k}}$.
Here, $\{\pi(s_{j_{l}}\cdots s_{j_{k}})\Omega\}_{l=1}^{k}$ is 
called the \textit{cycle} of $P(J)$.
\item
Let $\nset{\infty}\!\equiv\!\{\!(i_{n})_{n\in {\bf N}}\!:\!\mbox{for any } 
n,\,i_{n}\!\in\!\nset{}\!\}$.
For $\!J\!=\!(j_{n})_{n\in{\bf N}}\!\in\!\nset{\infty}$,
$P(J)$ is the class of representations $({\cal H}, \pi)$  of $\con$
with a cyclic unit vector $\Omega\in {\cal H}$ such that
$\!\{\pi((s_{J_{(n)}})^{*})\Omega\!:\!n\!\in\!{\bf N}\}$
is an orthonormal family in ${\cal H}$ where
$J_{(n)}\!\equiv\! (j_{1},\ldots,j_{n})$.
Here, 
$\!\{\pi((s_{J_{(n)}})^{*})\Omega\!:\!n\!\in\!{\bf N}\}$ is 
called the \textit{chain} of $P(J)$.
\end{enumerate}
\end{defi}

\noindent
We simply call that $({\cal H},\pi)$ is a {\it {\it cycle} ({\it chain})} if
there is $J\in\nset{k}$ ({\it resp.} $J\in\nset{\infty}$)
such that $({\cal H},\pi)$ belongs to $P(J)$.

Let $UHF_{N}$ be the uniformly hyperfinite algebra and
fix a generating set $\bigcup_{l\geq 1}\{E_{K,L}\}_{J,K\in\nset{l}}$ of $UHF_{N}$
such that $\{E_{K,L}\}_{J,K\in\nset{l}}$ is a system of matrix units of 
a unital subalgebra of $UHF_{N}$ which is isomorphic to $M_{N^{l}}({\bf C})$
and $E_{K,L}=\sum_{i=1}^{N}E_{K\cup (i),L\cup (i)}$
for each $K,L\in\nset{l}$ and $l\geq 1$
where $K\cup (i)\equiv (k_{1},\ldots,k_{l},i)$ with $K=(k_{1},\ldots,k_{l})$.
If there is no ambiguity, then $E_{K,L}$ is also denoted as $E_{KL}$.
The following definition depends on the choice of 
$\bigcup_{l\geq 1}\{E_{KL}\}_{J,K\in\nset{l}}$.
%
%
\begin{defi}[Permutative representation of $UHF_{N}$]
\label{defi:uhf}
Let $({\cal H},\pi)$ be a  representation of $UHF_{N}$.
\begin{enumerate}
\item
For $J=(j_{n})_{n\in {\bf N}}\in\nset{\infty}$,
$P[J]$ is the class of representations $({\cal H}, \pi)$ with 
a cyclic unit vector $\Omega\in {\cal H}$ such that
$\pi(E_{J_{n},J_{n}})\Omega=\Omega$ for each $n\geq 1$
where $J_{n}\equiv (j_{1},\ldots,j_{n})$.
\item
For $J=(j_{n})_{n=1}^{k}\in\nset{k}$,
$P[J]$ is the class of representations $({\cal H}, \pi)$ such that
$({\cal H},\pi)$ is $P[J^{\infty}]$.
\end{enumerate}
\end{defi}

\noindent
In both Definition \ref{defi:first} and \ref{defi:uhf},
we call $\Omega$ the {\it GP (Generalized Permutative) vector} 
of $({\cal H},\pi)$.

Identifying $UHF_{N}$ with $\con^{U(1)}$ by the embedding 
$\Phi_{UHF_{N}}$ of $UHF_{N}$ into $\con$ as
%
%
\begin{equation}
\label{eqn:mapone}
\Phi_{UHF_{N}}(E_{JK})\equiv s_{J}s_{K}^{*}\quad (J,K\in\nset{n},\, n\geq 1),
\end{equation}

\noindent
we show explicit \brl s of permutative representations of $\con$
restricted on $UHF_{N}$.
In the case that a representation $({\cal H},\pi)$ of $\con$ is $P(J)$,
we denote the restriction $({\cal H},\pi|_{UHF_{N}})$ by 
$P(J)|_{UHF_{N}}$ for simplicity of description.
%
%
\begin{Thm}
\label{Thm:restriction}
For $J=(j_{1},\ldots,j_{l})\in\nset{l}$ 
and $K=(k_{n})_{n\in {\bf N}}\in\nset{\infty}$,
the following irreducible decompositions hold:
%
%
\begin{equation}
P(J)|_{UHF_{N}}=\bigoplus_{\sigma\in {\bf Z}_{l}}P[\sigma J],\quad
P(K)|_{UHF_{N}}=\bigoplus_{\eta\in {\bf Z}}P[\eta K], 
\end{equation}
where $\sigma J\equiv (j_{\sigma(1)},\ldots,j_{\sigma(l)})$
and $\eta K=(k_{n}^{'})_{n\in {\bf N}}$ is defined by
$k_{n}^{'}\equiv k_{\eta(n)}$ 
for $\eta(n)\geq 1$, $k_{n}^{'}\equiv 1$ for $\eta(n)\leq 0$.
\end{Thm}

\noindent
In \cite{BJ}, this result is described in terms of 
the ``atom" of a permutative representation.
The decomposition of the chain is a special case of \cite{BC}.

Let ${\goth S}_{N,l}$ be the set of all permutations on the set $\nset{l}$.
For $\sigma\in {\goth S}_{N,l}$,
define the endomorphism $\psi_{\sigma}$ of $\con$  by
%
%
\begin{equation}
\label{eqn:defeqn}
\psi_{\sigma}(s_{i})\equiv 
u_{\sigma}s_{i}\quad(i\edot),
\end{equation}
where $u_{\sigma}\equiv \sum_{J\in \nset{l}}s_{\sigma(J)}(s_{J})^{*}$.
It should be noted
that $\psi_{\sigma}$ commutes with the $U(1)$-gauge action on $\con$
for any $\sigma$.
In \cite{PE01}, it is shown the following:
Numbers of unitary equivalence classes of elements in 
$E_{2,2}\equiv \{\psi_{\sigma}:\sigma\in {\goth S}_{2,2}\}$ is $16$.
The group $G_{2}\equiv {\rm Aut}\co{2}\cap E_{2,2}$ 
is isomorphic to the Klein's four-group, which
consists of two outer and two inner automorphisms.
The set $E_{2,2}\setminus G_{2}$
consists of $10$ irreducible and $10$ reducible endomorphisms
and numbers of their equivalence classes are $5$ and $9$, respectively.
Especially, (the class of) the canonical endomorphism
of $\co{2}$ belongs to the set of reducible classes in $E_{2,2}$.
Now, we consider the restriction of $E_{2,2}$ to $UHF_{2}$.
Since $\rho|_{UHF_{2}}$ is also an endomorphism of $UHF_{2}$
for each $\rho\in E_{2,2}$, its properties are obtained as follows:

%
%
\begin{Thm}
\label{Thm:mainfour}
Define $UE_{2,2}\equiv \{\rho|_{UHF_{2}}:\rho\in E_{2,2}\}$.
\begin{enumerate}
\item
The cardinarity of $UE_{2,2}$ equals $20$ and
the number of unitary equivalence classes of elements in $UE_{2,2}$ is $12$.
%
\item
The subgroup $UG_{2}\equiv {\rm Aut}UHF_{2}\cap UE_{2,2}$ of ${\rm Aut}UHF_{2}$
is also isomorphic to the Klein's four-group, which
consists of two outer and two inner automorphisms of $UHF_{2}$.
\item
The set of all equivalence classes in $UE_{2,2}\setminus UG_{2}$ consists of
$4$ irreducibles and $6$ reducibles.
\end{enumerate}
\end{Thm}

With the embedding $\Phi_{CAR}$ of the CAR algebra ${\cal A}$
into $\co{2}$ defined as (\ref{eqn:recursive}), and with the identification
${\cal A}$ and $UHF_{2}$ by a map $\Psi$ from ${\cal A}$ onto $UHF_{2}$,
%
%
\begin{equation}
\label{eqn:psi}
\Psi(a_{1})\equiv E_{12},\quad
\Psi(a_{n})\equiv \sum_{J\in\{1,2\}^{n-1}}(-1)^{n_{2}(J)}
E_{J\cup (1),J\cup (2)}\quad(n\geq 2),
\end{equation}
where $n_{2}(J)\equiv \sum_{i=1}^{k}(j_{i}-1)$ for 
$J=(j_{1},\ldots,j_{k})$,
we obtain a relation $\Phi_{CAR}= \Phi_{UHF_{2}}\circ \Psi$
among $\Psi$, $\Phi_{CAR}$ and 
$\Phi_{UHF_{2}}$ in (\ref{eqn:mapone}). 
Hence, according to these maps, we may regard ${\cal A}$ and $UHF_{2}$ as the
same subalgebra of $\co{2}$.
Restrictions of representations and branching laws
are also described by such identifications.

As an application of branching laws on $UHF_{2}$, we show the following:
Let $({\cal H},\pi)$ be the Fock representation
of ${\cal A}$ with the vacuum $\Omega$.
We identify ${\cal A}$ with $\pi({\cal A})$.
Then, we have $a_{n}\Omega=0$ for each $n\geq 1$.
Let ${\bf Z}+1/2\equiv \{n+1/2:n\in {\bf Z}$ and
${\bf Z}_{\geq }+1/2\equiv \{x\in {\bf Z}+1/2: x> 0\}$.
Define a mixture $\{b_{k}\}_{k\in {\bf Z}+1/2}$ of fermions by
%
%
\begin{equation}
\label{eqn:composition}
\left\{
\begin{array}{ll}
b_{k}\equiv&(-1)^{k-1/2}(a_{1}a_{1}^{*}a_{2k+2}^{*}+a_{1}^{*}a_{1}a_{2k+2}),\\
\\
b_{-k}\equiv&(-1)^{k-1/2}(a_{1}a_{1}^{*}a_{2k+1}-a_{1}^{*}a_{1}a_{2k+1}^{*})\\
\end{array}
\right.
\quad(k\in {\bf Z}_{\geq}+1/2).
\end{equation}
Then, we obtain $b_{k}b_{l}+b_{l}b_{k}=0$,  
$b_{k}b_{l}^{*}+b_{l}^{*}b_{k}=\delta_{kl}I$ for each $k,l$, and
%
%
\begin{equation}
b_{k}\Omega=(-1)^{k-1/2}a_{2k+2}^{*}\Omega,\quad
b_{-k}^{*}\Omega=(-1)^{k-1/2}a_{2k+1}^{*}\Omega,\quad
b_{k}^{*}\Omega=b_{-k}\Omega=0
\end{equation}
for $k>0$.
It is shown that the vectors
%
%
\begin{equation}
b_{k_{1}}\cdots b_{k_{n}}b_{-l_{1}}^{*}\cdots b_{-l_{m}}^{*}\Omega
\quad(k_{1},\ldots,k_{n},l_{1},\ldots,l_{m}\in {\bf Z}_{\geq}+1/2)
\end{equation}
span the infinite wedge representation of ${\cal A}$ with 
the Dirac vacuum (or the vacuum of the infinite wedge) $\Omega$ \cite{MJD,Oko01}.
In this way,  it becomes possible to reinterpret   
the Fock vacuum as the Dirac vacuum. 
On the other hand, for $\Omega^{*}\equiv a_{1}^{*}\Omega$, the following 
equations hold:
%
%
\begin{equation}
b_{-k}\Omega^{*}\!=\!(-1)^{k-1/2}a_{2k+1}^{*}\Omega^{*},\ \ 
b_{k}^{*}\Omega^{*}\!=\!(-1)^{k-1/2}a_{2k+2}^{*}\Omega^{*},\ \ 
b_{k}\Omega^{*}\!=\!b_{-k}^{*}\Omega^{*}=0.
\end{equation}
Hence, the vectors
%
%
\begin{equation}
b_{-k_{1}}\cdots b_{-k_{n}}b_{l_{1}}^{*}\cdots b_{l_{m}}^{*}\Omega^{*}
\quad(k_{1},\ldots,k_{n},l_{1},\ldots,l_{m}\in {\bf Z}_{\geq}+1/2)
\end{equation}
span the dual infinite wedge representation of ${\cal A}$ 
with the dual Dirac vacuum $\Omega^{*}$.
In consequence, 
through the mixture (\ref{eqn:composition}) of fermions,
the Fock representation is transformed to the direct sum of 
the infinite wedge representation and its dual:
%
%
\begin{equation}
Fock\quad \stackrel{mixture}{\Rightarrow}\quad \mbox{{\it Infinite Wedge}}
\oplus \mbox{{\it Dual Infinite Wedge}}.
\end{equation}


The present paper is organized as follows:
We show properties of permutative representations and permutative endomorphisms 
of the Cuntz algebras and the uniformly hyperfinite algebras 
in $\S$ \ref{section:second}.
In $\S$ \ref{subsection:secondtwo}, Theorem \ref{Thm:restriction} is proved.
We treat the second order permutative endomorphisms and their branching laws
in $\S$ \ref{section:third}.
In $\S$ \ref{subsection:thirdtwo}, Theorem \ref{Thm:mainfour} is proved.
We apply these results to the case of fermions in $\S$ \ref{section:fourth}. 
In $\S$ \ref{subsection:fourthtwo}, 
we explain (\ref{eqn:composition}) and its meaning in the representation theory.
%
%
\sftt{Branching laws on $\con$ and $UHF_{N}$}
\label{section:second}

In this paper, any representation and endomorphism
are assumed unital and $*$-preserving.

%
%
\ssft{On $\con$}
\label{subsection:secondone}
For $N\geq 2$, let $\con$ be the {\it Cuntz algebra} \cite{C}, that is,
a C$^{*}$-algebra which is universally generated by
generators $s_{1},\ldots,s_{N}$ satisfying
$s_{i}^{*}s_{j}=\delta_{ij}I$ for $i,j\edot$ and
$s_{1}s_{1}^{*}+\cdots +s_{N}s_{N}^{*}=I$.

We review results of permutative representations \cite{BJ,DaPi2,DaPi3}.
$({\cal H},\pi)$ is a {\it permutative representation} of $\con$
if there is a complete orthonormal basis $\{e_{n}\}_{n\in\Lambda}$
of ${\cal H}$ and a family $f=\{f_{i}\}_{i=1}^{N}$
of maps on $\Lambda$ such that $\pi(s_{i})e_{n}=e_{f_{i}(n)}$
for each $n\in\Lambda$ and $i\edot$.
Any permutative representation is uniquely decomposed into
cyclic permutative representations up to unitary equivalence.
For any $J$, $P(J)$ contains only one unitary equivalence class.
Any cyclic permutative representation is equivalent to $P(J)$ for a certain 
$J\in \nset{\#}\equiv\coprod_{k\geq 1}\nset{k}\sqcup \nset{\infty}$.

We prepare several notions of multiintegers.
Define $\nset{*}_{1}\equiv\coprod_{k\geq 1}\nset{k}$ and
$\nset{*}\equiv\coprod_{k\geq 0}\nset{k}$, $\nset{0}\equiv \{0\}$.
The {\it length} $|J|$ of $J\in \nset{\#}$ is defined by
$|J|\equiv k$ for $J\in \nset{k}$.
For $J_{1},J_{2}\in\nset{*}$ and $J_{3}\in\nset{\infty}$,
$J_{1}\cup J_{2}\equiv(j_{1},\ldots,j_{k},j_{1}^{'},\ldots,j_{l}^{'})$,
$J_{1}\cup J_{3}\equiv(j_{1},\ldots,j_{k},j_{1}^{''},j_{2}^{''},\ldots)$
for $J_{1}=(j_{a})_{a=1}^{k}$, $J_{2}=(j_{b}^{'})_{b=1}^{l}$
and $J_{3}=(j_{n}^{''})_{n\in {\bf N}}$.
Especially, we define $J\cup (0)=(0)\cup J=J$ for convention.
For $J\in\nset{*}$ and $k\geq 2$,
$J^{k}\equiv J\cup\cdots\cup J$ ($k$ times).
$J\in\nset{*}_{1}$ is {\it periodic} if there exist an integer $m\geq 2$
and a multiintegers $J_{0}\in \nset{*}_{1}$ such that $J=(J_{0})^{m}$.
For $J_{1},J_{2}\in \nset{*}_{1}$, $J_{1}\sim J_{2}$
if $J_{1},J_{2}\in\nset{k}$ ($k\geq 1$) and $J_{2}
=(j_{p},\ldots,j_{k},j_{1},\ldots,j_{p-1})$ ($1\leq p \leq k$)
with $J_{1}\equiv(j_{1},\ldots,j_{k})$.
For $(J,z), (J^{'},z^{'})\in \nset{*}_{1}\times U(1)$, 
$(J,z)\sim(J^{'},z^{'})$ if $J\sim J^{'}$ and $z=z^{'}$.
For $J_{1}=(j_{l})_{l=1}^{k}, J_{2}=(j^{'}_{l})_{l=1}^{k}$, 
$J_{1}\prec J_{2}$ if $\sum_{l=1}^{k}(j_{l}^{'}-j_{l})N^{k-l}\geq 0$.
Especially, any element in $\nset{}$ is nonperiodic.
$J\in\nset{\infty}$ is {\it \evp} if there are $J_{0},J_{1}\in\nset{*}_{1}$
such that $J=J_{0}\cup (J_{1})^{\infty}$.
For $J_{1},J_{2}\in\nset{\infty}$, $J_{1}\sim J_{2}$
if there exist $J_{3},J_{4}\in\nset{*}$ and $J_{5}\in\nset{\infty}$ 
such that $J_{1}=J_{3}\cup J_{5}$ and $J_{2}=J_{4}\cup J_{5}$.

Next, we introduce a representation of $\con$ which is not a permutative one.
%
%
\begin{defi}
\label{defi:firsttwo}
$GP(\pm)$ is the class of representations $({\cal H}, \pi)$ with
 a cyclic unit vector $\Omega\in {\cal H}$
such that $\pi(s_{1}\pm s_{2})\Omega=\sqrt{2}\Omega$.
\end{defi}

\noindent
For $({\cal H},\pi)$ of $P(J)$ and $\rho\in {\rm End}\con$,
we denote $({\cal H},\pi\circ \rho)$ by $P(J)\circ \rho$ 
for simplicity of description.
%
%
\begin{Thm}
\label{Thm:permurep}
Define $P(J;z^{k})\equiv P(J)\circ \gamma_{z}$ for $z\in U(1)$ 
and $J\in\nset{k}$.
\begin{enumerate}
\item
For $(J,z)\in \nset{*}_{1}\times U(1)$, 
$P(J;z)$ is irreducible if and only if $J$ is nonperiodic.
For $J\in \nset{\infty}$, $P(J)$ is irreducible if and only if $J$ is not \evp.
\item
For $(J_{1},z_{1}),(J_{2},z_{2})\in \nset{*}_{1}\times U(1)$,
$P(J_{1};z_{1})\sim P(J_{2};z_{2})$ if and only if 
$(J_{1},z_{1})\sim (J_{2},z_{2})$.
For $J_{1},J_{2}\in \nset{\infty}$, $P(J_{1})\sim P(J_{2})$
if and only if $J_{1}\sim J_{2}$.
\item
For $J\in\nset{*}_{1}$ and $l\geq 1$,
$P(J^{l})=P(J;\xi_{1})\oplus\cdots\oplus P(J;\xi_{l})$
where $\xi_{n}\equiv e^{2\pi\sqrt{-1}(n-1)/l}$.
This decomposition is unique up to unitary equivalence.
Especially, if $J$ is nonperiodic, then this is multiplicity-free.
\item
$GP(\pm)$ contains only one unitary equivalence class and it is irreducible.
Furthermore, $GP(+)$ and $GP(-)$ are not equivalent.
\end{enumerate}
\end{Thm}
%
%
\pr
(i)-(iii) are proved in \cite{BJ,DaPi2,DaPi3,K1}.

\noindent (iv)
Define $\phi\in {\rm Aut}\con$ by
$\phi(s_{1})\equiv (s_{1}+ s_{2})/\sqrt{2}$,
$\phi(s_{2})\equiv (s_{1}- s_{2})/\sqrt{2}$
and $\phi(s_{i})\equiv s_{i}$ for each $i=3,\ldots,N$.
For any representation $({\cal H},\pi)$, 
$(\pi\circ \phi^{-1})(s_{1}+ s_{2})=\sqrt{2}\pi(s_{1})$
and
$(\pi\circ \phi^{-1})(s_{1}- s_{2})=\sqrt{2}\pi(s_{2})$.
This implies that 
$P(1)\circ \phi^{-1}=GP(+)$ and $P(2)\circ \phi^{-1}=GP(-)$.
Therefore, from (ii) and (iii), the statements are verified.
\qedh

Hereafter, we denote a representation $({\cal H},\pi)$ of $\con$ by $\pi$
for simplicity of description.
%
%
\begin{Thm}
\label{Thm:mainone}
For $\psi_{\sigma}$ in (\ref{eqn:defeqn}), the following holds:
\begin{enumerate}	
\item
If $\pi$ is a permutative representation,
then $\pi\circ \psi_{\sigma}$ is also a permutative representation.
\item
If $\pi$ is $P(J)$ for $J\in\nset{\#}$ and $\sigma\in {\goth S}_{N,l}$,
then there exist multiintegers $J_{1},\ldots,J_{M}\in\nset{\#}$ and
subrepresentations $\pi_{1},\ldots,\pi_{M}$
of $\pi\circ \psi_{\sigma}$ such that
%
%
\begin{equation}
\label{eqn:decoeqone}
\pi\circ \psi_{\sigma}=\pi_{1}\oplus\cdots\oplus\pi_{M}
\end{equation}
with $\pi_{i}$ being $P(J_{i})$ for $i=1,\ldots,M$,
and $1\leq M\leq N^{l-1}$.
\item
In (ii),
if $J\!\in\!\nset{k}\!$ ($J\!\in\!\nset{\infty}\!$),
then $J_{i}\!\in\!\coprod_{a=1}^{N^{l-1}}\!\nset{ak}$
({\it resp}. $J_{i}\!\in\!\nset{\infty}$) for $i=1,\ldots,M$.
\end{enumerate}
\end{Thm}
%
%
\pr
See Theorem 1.3 of \cite{PE02}. 
\qedh

\noindent
The endomorphism
$\psi_{\sigma}$ in (\ref{eqn:defeqn}) is called the {\it permutative
endomorphism} of $\con$ associated with $\sigma$.

From the uniqueness of decomposition of the
permutative representation,
the rhs in (\ref{eqn:decoeqone}) is unique up to unitary equivalence.
Then (\ref{eqn:decoeqone}) can be rewritten as follows:
%
%
\begin{equation}
\label{eqn:simpdeco}
P(J)\circ \psi_{\sigma}=P(J_{1})\oplus\cdots\oplus P(J_{M}).
\end{equation}
We call (\ref{eqn:simpdeco}) the {\it \brl} for $\psi_{\sigma}$ 
with respect to $P(J)$.
The  \brl\ for $\psi_{\sigma}$ is unique up to
unitary equivalence of $\psi_{\sigma}$.
From contraposition to this result,
we distinguish two equivalence classes of endomorphisms
in $\S$\ref{section:third}.

%
%
\ssft{On $UHF_{N}$}
\label{subsection:secondtwo}
%
%
\begin{lem}
\label{lem:state}
For $J\in\nset{\infty}$, let $\omega$ be the state of $UHF_{N}$ defined by
%
%
\begin{equation}
\label{eqn:state}
\omega(E_{K,L})\equiv 0\quad(K\ne L),\quad
\omega(E_{K,K})\equiv \delta_{K,J_{n}}\quad(|K|=n).
\end{equation}
Then the GNS representation of $UHF_{N}$ by $\omega$ is $P[J]$ 
in Definition \ref{defi:uhf}.
\end{lem}
%
%
\pr
Let $\pi$ be $P[J]$ with the GP vector $\Omega$.
Define the state $\rho$ of $UHF_{N}$ by $\rho\equiv <\Omega|\pi(\cdot)\Omega>$.
For $\omega$ in (\ref{eqn:state}),
we can verify that $\rho=\omega$.
Because of the cyclicity of $\Omega$
and the uniqueness of the GNS representation, the statement holds.
\qedh

\noindent
For $J=(j_{n})_{n\in {\bf N}},J^{'}=(j^{'}_{n})_{n\in {\bf N}}\in\nset{\infty}$,
$J\approx J^{'}$ if there exists an integer $n_{0}\geq 1$ 
such that $j_{n}=j^{'}_{n}$ for each $n\geq n_{0}$.
%
%
\begin{prop}
\label{prop:uhf}
\hfill
\begin{enumerate}
\item
For any $J\in\nset{\#}$, 
$P[J]$ contains only one unitary equivalence class.
\item
For any $J\in\nset{\#}$, $P[J]$ is irreducible.
\item
For $J,J^{'}\in\nset{\infty}$, $P[J]\sim P[J^{'}]$ if and only 
if $J\approx J^{'}$.
\item
For $J,J^{'}\in\nset{*}_{1}$, $P[J]\sim P[J^{'}]$ if and only if $J=J^{'}$.
\end{enumerate}
\end{prop}
%
%
\pr
(i) By Lemma \ref{lem:state},  the statement holds.

\noindent
(ii)
Let $\vep_{1},\ldots,\vep_{N}$ be the standard basis of ${\bf C}^{N}$.
Then, we see that the state $\omega$ in (\ref{eqn:state}) equals to the state 
$F_{(0)}^{\phi}$ in Definition 2.3 of \cite{BC} 
for $\phi=(\vep_{j_{n}})_{n\in {\bf N}}$.
Because $F_{(0)}^{\phi}$ is pure, the statement holds.

\noindent (iii)
Applying Theorem 2.5 in \cite{BC} to $P[J]$, the assertion holds.

\noindent (iv)
From (iii) and the definition of $P[J]$,
we have $P[J]\sim P[J^{'}]$ if and only if $J^{\infty}\approx (J^{'})^{\infty}$.
This is equivalent  with $J=J^{'}$.
\qedh


\noindent
{\bf Proof of Theorem \ref{Thm:restriction}.}
Let $({\cal H},\pi)$ be $P(J)$ of $\con$ with the GP vector $\Omega$.
Here we denote $\pi(s_{i})$ by $s_{i}$ for simplicity of description.
In both decomposition formulae in the statement, 
the irreducibility of each component follows by Proposition \ref{prop:uhf} (ii).

Assume that $J=(j_{1},\ldots,j_{l})$.
From $UHF_{N}=\con^{U(1)}$, we obtain
$P(J;z)|_{UHF_{N}}=P(J)|_{UHF_{N}}$ for any $z\in U(1)$.
Hence, it is sufficient to show that 
$P(J)|_{UHF_{N}}=\bigoplus_{\sigma\in {\bf Z}_{l}}P[\sigma J]$.
Define $V_{i}$ by the completion of $V_{i,0}\equiv\pi(UHF_{N})e_{i}$ 
and $e_{i}\equiv s_{j_{i}}\cdots s_{j_{l}}\Omega$ for $i=1,\ldots,l$.
Now, we show that ${\cal H}=V_{1}\oplus\cdots\oplus V_{l}$.
By setting $E_{LM}\equiv s_{L}s_{M}^{*}$ for $L,M\in\nset{a}$,
we see that if $i\ne j$, then $<E_{LM}e_{i}|E_{L^{'}M^{'}}e_{j}>=0$ 
for each $L,M\in\nset{a}$ and $L^{'},M^{'}\in\nset{b}$.
Therefore $V_{i}$ and $V_{j}$ are orthogonal for $i\ne j$.
On the other hand,
if $|L|=ln+i$ for $n\geq 0$ and $i=0,1,\ldots,l-1$, then
$s_{L}\Omega=s_{L}(s_{j_{l-i+1}}\cdots s_{j_{l}}(s_{J})^{n})^{*}e_{l-i+1}$ 
and $s_{L}(s_{j_{l-i+1}}\cdots s_{j_{l}}(s_{J})^{n})^{*}\in UHF_{N}$. 
Hence, we have $s_{L}\Omega\in V_{i}$.
Because ${\rm Lin}\langle\{s_{L}\Omega:L\in\nset{*}\}\rangle$ 
is dense in ${\cal H}$, 
we have ${\cal H}=V_{1}\oplus\cdots\oplus V_{l}$ as a $UHF_{N}$-module.
Define a state $\omega_{i}\equiv <e_{i}|\pi(\cdot)e_{i}>$ 
of $UHF_{N}$ for $i=1,\ldots,l$
and $\sigma_{i}J\equiv (j_{i},\ldots,j_{l},j_{1},\ldots,j_{i-1})$.
Then $\omega_{i}$ satisfies (\ref{eqn:state})
with respect to $(\sigma_{i} J)^{\infty}$.
Because the restriction $\pi^{[i]}$ of $\pi|_{UHF_{N}}$ on $V_{i}$
is equivalent to the GNS representation by $\omega_{i}$,
we have $\pi^{[i]}$ is $P[\sigma_{i}J]$.
Therefore, the first formula holds.

Assume that $K=(k_{n})_{n\in {\bf N}}$.
Define $e_{n}\equiv s_{k_{n}}^{*}\cdots s_{k_{1}}^{*}\Omega$,
$e_{-n}\equiv s_{1}^{n}\Omega$ for $n\geq 1$ and $e_{0}\equiv \Omega$.
Define $V_{n}$ by the completion of $V_{n,0}\equiv\pi(UHF_{N})e_{n}$ 
for $n\in {\bf Z}$.
Then $V_{n}$ and $V_{m}$ are orthogonal for $n\ne m$.
Let $L\in\nset{a}$ and $n\in {\bf Z}$.
For $n\geq l$, we have
$s_{L}e_{n}=s_{L}s_{k_{n}}^{*}\cdots s_{k_{n-a+1}}^{*}e_{n-a}\in V_{n-a}$.
For $1\leq n< a$,
we have $s_{L}e_{n}=s_{L}s_{K_{n}}^{*}(s_{1}^{*})^{a-n}e_{n-a}\in V_{n-a}$.
For $n\leq 0$, 
we have $s_{L}e_{n}=s_{L}(s_{1}^{*})^{a}e_{n-a}\in V_{n-a}$.
Hence, we obtain that
$s_{L}e_{n}\in \bigoplus_{m\in {\bf Z}}V_{m}$ for each $n$ and $L$.
Because ${\rm Lin}\langle\{s_{L}e_{n}:L\in\nset{*},n\in {\bf Z}\}\rangle$
is dense in ${\cal H}$,
we have ${\cal H}=\bigoplus_{n\in {\bf Z}}V_{n}$.
Define a state $\omega_{n}$ of $UHF_{N}$ by
$\omega_{n}\equiv <e_{n}|\pi(\cdot)e_{n}>$, and 
$\sigma_{n}(m)\equiv m+n$ for $n,m\in {\bf Z}$.
Then $\omega_{n}$ satisfies (\ref{eqn:state}) with respect to $\sigma_{n}K$.
Likewise in the case of $J$,  the second formula holds.
\qedh

In order to classify endomorphisms of $UHF_{N}$,
we introduce two representations as follows:
%
%
\begin{defi}
\label{defi:uhftwo}
$GP[\pm]$ is the class of representations $({\cal H}, \pi)$ of $UHF_{N}$
with a cyclic vector $\Omega$ 
such that $\pi(F_{n,\pm})\Omega=\Omega$ for each $n\geq 1$,
where $F_{n,\pm}\in UHF_{N}$ is defined by
%
%
\begin{equation}
F_{n,+}\equiv 2^{-n}\sum_{J,K\in\{1,2\}^{n}}E_{JK},\quad
F_{n,-}\equiv 2^{-n}\sum_{J,K\in\{1,2\}^{n}}(-1)^{\|J-K\|}E_{JK}
\end{equation}
with $\|J-K\|\equiv \sum_{i=1}^{n}(j_{i}-k_{i})$
for $J=(j_{i})_{i=1}^{n}$ and $K=(k_{i})_{i=1}^{n}$.
We call $\Omega$ the GP vector of $({\cal H},\pi)$.
\end{defi}
%
%
\begin{prop}
\label{prop:restriction}
\hfill
\begin{enumerate}
\item
If $J$ is nonperiodic (not \evp), then the case (i)
(respectively (ii)) in Theorem \ref{Thm:restriction} is multiplicity-free.
\item $GP(\pm)|_{UHF_{N}}=GP[\pm]$.
\item
$GP[\pm]$ contains only one unitary equivalence class
and it is irreducible. $GP[+]$ and $GP[-]$ are not equivalent.
\end{enumerate}
\end{prop}
%
%
\pr
(i) This holds as a result of Proposition \ref{prop:uhf} (iii) and (iv).

\noindent (ii)
In the proof of Theorem \ref{Thm:permurep} (v),
we see that $\phi|_{UHF_{N}}\in {\rm Aut}UHF_{N}$ 
and $P[1]\circ \phi^{-1}=GP[+]$.
Hence,  we have
%
%
\begin{equation}
GP[+]\circ \phi=P[1]=P(1)|_{UHF_{N}}=(GP(+)\circ \phi)|_{UHF_{N}}.
\end{equation}
From this, we obtain $GP[+]=GP(+)|_{UHF_{N}}$.
In a similar way, we obtain $GP[-]=GP(-)|_{UHF_{N}}$.

\noindent (iii)
Because we have $P[1]\circ \phi^{-1}=GP[+]$ and $P[2]\circ \phi^{-1}=GP[-]$,
the statements hold.
\qedh

%
%
\sftt{Second order permutative endomorphisms}
\label{section:third}

In order to classify endomorphisms of C$^{*}$-algebras,
we prepare several notions for their properties.
Let $\enda$ be the set of all unital $*$-endomorphisms of 
a unital $*$-algebra ${\cal A}$ and $\rho,\rho_{1},\rho_{2}\in\enda$.
We state that  
$\rho$ is {\it proper} if $\rho({\cal A})\ne {\cal A}$;
$\rho$ is {\it irreducible} if $\rho({\cal A})^{'}\cap  {\cal A}={\bf C}I$;
$\rho$ is {\it reducible} if $\rho$ is not irreducible;
$\rho_{1}$ and $\rho_{2}$ are {\it equivalent} ($\rho_{1}\sim\rho_{2}$) 
if there exists a unitary
$u\in{\cal A}$ such that $\rho_{2}={\rm Ad}u\circ \rho_{1}$.

Then the following holds.
If $\rho_{1}$ is proper and $\rho_{2}$ is not, then $\rho_{1}\not\sim\rho_{2}$.
If $\rho_{1}$ is irreducible and $\rho_{2}$ is not, then
$\rho_{1}\not\sim\rho_{2}$.
Any automorphism is irreducible and not proper.
If ${\cal A}$ is simple, then $\rho$ is an automorphism
if and only if $\rho$ is not proper.
Let ${\rm Rep}{\cal A}$ be the class of all
unital $*$-representations of ${\cal A}$.
If ${\cal A}$ is simple and there is $\pi\in {\rm Rep}{\cal A}$
such that both $\pi$ and $\pi\circ \rho$ are irreducible, 
then $\rho$ is irreducible.
If there exists $\pi\in {\rm Rep}{\cal A}$ such that
$\pi\circ\rho_{1}\not\sim\pi\circ\rho_{2}$, then $\rho_{1}\not\sim\rho_{2}$.
If there exists $\pi\in {\rm Rep}{\cal A}$ such that $\pi$ is irreducible
and $\pi\circ\rho$ is not, then $\rho$ is proper.
If ${\cal B}$ is a subalgebra of ${\cal A}$
and $\rho\in {\rm End}{\cal A}$ satisfies
$\rho|_{{\cal B}}\in {\rm End}{\cal B}$, 
then $(\pi\circ \rho)|_{{\cal B}}=(\pi|_{{\cal B}})\circ (\rho|_{{\cal B}})$ 
for any $\pi\in {\rm Rep}{\cal A}$.

%
%
\subsection{On $\co{2}$}
\label{subsection:thirdone}
In order to derive branching laws on $UHF_{2}$,
we review the corresponding results on $\co{2}$.
Recall $\psi_{\sigma}$ for $\sigma\in {\goth S}_{N,l}$ in (\ref{eqn:defeqn}).
For $l=2$,
we call $\psi_{\sigma}$ the {\it second order permutative endomorphisms} 
of $\con$ by $\sigma$.
In \cite{PE01}, we show the complete classification of 
unitary equivalence classes of the second order permutative endomorphisms
of $\co{2}$ by using their branching laws.
By using a map $\lambda:\{1,2,3,4\}\to\{(1,1),(1,2),(2,1),(2,2)\}$ defined by
$\lambda(1)= (1,1)$, $\lambda(2)= (1,2)$,
$\lambda(3)= (2,1)$, $\lambda(4)= (2,2)$,
we identify $\sigma\in {\goth S}_{2,2}$
and $\lambda^{-1} \circ \sigma\circ\lambda\in {\goth S}_{4}$,
and use a notation
%
%
\begin{equation}
E_{2,2}=\{\psi_{\sigma}:\sigma\in{\goth S}_{4}(\cong {\goth S}_{2,2})\}.
\end{equation}
Their properties are summarized in Table 1 (see Table I in \cite{PE01}).
%
{\footnotesize
\[
\begin{array}{ccccc}
\multicolumn{5}{c}{\mbox{Table 1. Elements in $E_{2,2}$.}}\\
\hline
\psi_{\sigma}& \psi_{\sigma}(s_{1})&
\psi_{\sigma}(s_{2})&\mbox{{\rm property}}
& {\rm Ad}u\circ \psi_{\sigma}\\
\hline
\psi_{id} & s_{1} & s_{2}&
inn.aut&\psi_{(14)(23)}\\
\psi_{12}   & s_{12,1}+s_{11,2}& s_{2}& 
irr.end&\psi_{1324}\\
\psi_{13}   &s_{21,1}+s_{12,2}& s_{11,1}+s_{22,2}
& irr.end&\psi_{1432}\\
\psi_{14}   & s_{22,1}+s_{12,2}& s_{21,1}+s_{11,2}&
red.end&\psi_{14}\\
\psi_{23}   &s_{11,1}+s_{21,2}& s_{12,1}+s_{22,2}&
red.end&\psi_{23}\\
\psi_{24}   &s_{11,1}+s_{22,2}&s_{21,1}+s_{12,2}&
irr.end&\psi_{1234}\\
\psi_{34}   &s_{1}& s_{22,1}+s_{21,2}&
irr.end&\psi_{1423}\\
\psi_{123} &s_{12,1}+s_{21,2}& s_{11,1}+s_{22,2}&
red.end&\psi_{243}\\
\psi_{132} &s_{21,1}+s_{11,2}& s_{12,1}+s_{22,2}&
red.end&\psi_{132}\\
\psi_{124} &s_{12,1}+s_{22,2}& s_{21,1}+s_{11,2}&
red.end&\psi_{124}\\
\psi_{142} &s_{22,1}+s_{11,2}& s_{21,1}+s_{12,2}&
irr.end&\psi_{134}\\
\psi_{134} &s_{21,1}+s_{12,2}& s_{22,1}+s_{11,2}&
irr.end&\psi_{142}\\
\psi_{143} &s_{22,1}+s_{12,2}& s_{11,1}+s_{21,2}&
red.end&\psi_{143}\\
\psi_{234} & s_{11,1}+s_{21,2}& s_{22,1}+s_{12,2}&
red.end&\psi_{234}\\
\psi_{243} &s_{11,1}+s_{22,2}& s_{12,1}+s_{21,2}&
red.end&\psi_{123}\\
\psi_{1234} &s_{12,1}+s_{21,2}& s_{22,1}+s_{11,2}&
irr.end &\psi_{24} \\
\psi_{1243} &s_{12,1}+s_{22,2}& s_{11,1}+s_{21,2}&
red.end&\psi_{1243}\\
\psi_{1324} &s_{2}& s_{12,1}+s_{11,2}&
irr.end&\psi_{12}\\
\psi_{1342} &s_{21,1}+s_{11,2}& s_{22,1}+s_{12,2}&
red.end&\psi_{1342}\\
\psi_{1423} &s_{22,1}+s_{21,2}& s_{1}&
irr.end&\psi_{34}\\
\psi_{1432} &s_{22,1}+s_{11,2}& s_{12,1}+s_{21,2}&
irr.end&\psi_{13}\\
\psi_{(12)(34)} &s_{12,1}+s_{11,2}& s_{22,1}+s_{21,2}&
out.aut&\psi_{(13)(24)}\\
\psi_{(13)(24)} &s_{2}& s_{1}&
out.aut &\psi_{(12)(34)}\\
\psi_{(14)(23)} &s_{22,1}+s_{21,2}& 
s_{12,1}+s_{11,2}& 
inn.aut&\psi_{id}\\
\botrule
\\
\multicolumn{5}{c}{
\begin{minipage}[c]{8.5cm}
The symbols are defined as 
$s_{ij,k}\equiv s_{i}s_{j}s_{k}^{*}$ for $i,j,k=1,2$, and
``inn.aut", ``out.aut", ``irr.end" and ``red.end" mean  
an inner automorphism, an outer automorphism, a proper irreducible endomorphism 
and a reducible endomorphism, respectively,
and $u\equiv s_{1}s_{2}^{*}+s_{2}s_{1}^{*}$.
\end{minipage}
}
\end{array}
\]
}

For $\zeta=(\zeta_{1},\zeta_{2}) \in\co{2}\times \co{2}$
and $\varphi_{1},\varphi_{2}\in {\rm End}\co{2}$,
define a linear $*$-preserving transformation
$\varphi_{1}+_{\zeta}\varphi_{2}$ on $\co{2}$ by
$\zeta_{1}\varphi_{1}(\cdot)\zeta_{1}^{*}
+\zeta_{2}\varphi_{2}(\cdot)\zeta_{2}^{*}$.
Let $\xi\equiv (s_{1},s_{2}),\ 
\xi^{'}\equiv ((s_{1}+s_{2})/\sqrt{2},\ (s_{1}-s_{2})/\sqrt{2})
\in\co{2}\times\co{2}$.
Then both $\varphi_{1}+_{\xi}\varphi_{2}$ and 
$\varphi_{1}+_{\xi^{'}}\varphi_{2}$ are endomorphisms of $\co{2}$
for any $\varphi_{1},\varphi_{2}$.
By using this notation, we can verify the following \cite{SE01}:
%
%
\begin{equation}
\label{eqn:sector}
\left\{
\begin{aligned}[c]
\psi_{14}&=\alpha+_{\xi^{'}}\alpha\theta,\\ 
\psi_{124}&=\alpha+_{\xi^{'}}\alpha\beta_{2},\\
\psi_{234}&=\iota +_{\xi^{'}}\beta_{2},
\end{aligned}
\quad
\begin{aligned}[c]
\psi_{23}&=\iota +_{\xi}\iota,\vphantom{+_{\xi^{'}}}\\
\psi_{132}&=\iota +_{\xi^{'}}\beta_{1},\\
\psi_{1243}&=\alpha+_{\xi}\alpha,\vphantom{+_{\xi^{'}}}
\end{aligned}
\quad
\begin{aligned}[c]
\psi_{123}&=\iota +_{\xi}\alpha,\vphantom{+_{\xi^{'}}}\\
\psi_{143}&=\alpha+_{\xi^{'}}\alpha\beta_{1},\\
\psi_{1342}&=\iota +_{\xi^{'}}\theta,
\end{aligned}
\right.
\end{equation}
where $\iota$ is the identity map on $\co{2}$,
$\alpha,\beta_{1},\beta_{2}\in {\rm Aut}\co{2}$ are defined by 
$\alpha(s_{i})\equiv s_{3-i}$,
$\beta_{j}(s_{i})\equiv (-1)^{\delta_{ij}}s_{i}$ for $i,j=1,2$
and $\theta\equiv \beta_{1}\circ\beta_{2}$.
On the other hand, the following holds from 
Definition \ref{defi:firsttwo} (ii):
%
%
\begin{equation}
\label{eqn:general}
GP(+)\circ \beta_{2}=GP(-),\quad GP(+)\circ \alpha=GP(+).
\end{equation}

\noindent
There are $16$ unitary equivalence classes in $E_{2,2}$ by \cite{PE02}.
We choose $16$ representatives in $E_{2,2}$ and show 
their branching laws in Table 2
by using Table II in \cite{PE01}, (\ref{eqn:sector}) and (\ref{eqn:general}).
%
%
{\footnotesize
\[
\begin{array}{@{}ccccc@{}}
\multicolumn{5}{c}{\mbox{Table 2. Branching laws for $E_{2,2}$ on $\co{2}$.}}\\
\toprule
\psi_{\sigma}& 
P(1)\circ\psi_{\sigma}
&
P(2)\circ\psi_{\sigma}
&
P(12)\circ\psi_{\sigma}
&
GP(+)\circ\psi_{\sigma}\\
\colrule
\psi_{id}      &P(1)&P(2)&P(12)&GP(+)\\
\psi_{(12)(34)}&P(2)&P(1)&P(12)&GP(+)\\
\psi_{12}&P(12)&P(1)\oplus P(2)&P(1122)&\mbox{---}\\
\psi_{13}&P(2) &P(2)           &P(11)  &\mbox{---}\\
\psi_{24}&P(1) &P(1)           &P(22)  &\mbox{---}\\
\psi_{34}&P(1)\oplus P(2)&P(12)&P(1122)&\mbox{---}\\
\psi_{142} &P(12)&P(12)&P(11)\oplus P(22)&\mbox{---}\\
\psi_{14}  &P(22)&P(11)&P(12)\oplus P(12)&\ GP(+)\oplus GP(+)\circ \theta\\
\psi_{23}  &P(1)\oplus P(1)&\ P(2)\oplus P(2)&\ P(12)\oplus P(12)&GP(+)\oplus GP(+)\\
\psi_{123}&P(1)\oplus P(2)&P(1)\oplus P(2)&P(12)\oplus P(12)&GP(+)\oplus GP(+)\\
\psi_{124}&P(22)&P(1)\oplus P(1)&P(1212)&GP(+)\oplus GP(-)\\
\psi_{132}&P(11)&P(2)\oplus P(2)&P(1212)&GP(+)\oplus GP(-)\circ \theta\\
\psi_{143}&P(2)\oplus P(2)&P(11)&P(1212)&GP(+)\oplus GP(-)\circ \theta\\
\psi_{234}&P(1)\oplus P(1)&P(22)&P(1212)&GP(+)\oplus GP(-)\\
\psi_{1243}&P(2)\oplus P(2)&P(1)\oplus P(1)&P(12)\oplus P(12)&GP(+)\oplus GP(+)\\
\psi_{1342}&P(11)&P(22)&P(12)\oplus P(12)&GP(+)\oplus GP(+)\circ \theta\\
\botrule
\\
\multicolumn{5}{c}{
\begin{minipage}[c]{11cm}
The part ``---" is omitted because it is complicated
and it is not necessary to classify $\psi_{\sigma}$'s in this paper.
\end{minipage}
}
\end{array}
\]
}
%
%
%
%
\ssft{On $UHF_{2}$}
\label{subsection:thirdtwo}
The restriction of each of $\psi_{\sigma}$'s in Table 1 on $UHF_{2}$ 
is also an endomorphism of $UHF_{2}$. 
We denote it by the same symbol for simplicity of description.
For representations $\pi$ and $\pi^{'}$ of a C$^{*}$-algebra ${\cal A}$,
we denote $\pi^{'}\prec \pi$
if $\pi^{'}$ is equivalent to a subrepresentation of $\pi$.
%
%
\begin{lem}
\label{lem:branchingone}
%
%
\begin{gather}
P[12]\circ \psi_{12}=P[1122]\oplus P[2211],\quad
P[21]\circ \psi_{12}=P[1221]\oplus P[2112], \notag\\
P[12]\circ \psi_{13}=P[21]\circ \psi_{13}=P[1]. 
\label{eqn:onethree}
\end{gather}
\end{lem}
%
%
\pr
By Table 2, we have $P(12)\circ \psi_{12}=P(1122)$.
On the other hand, by Theorem \ref{Thm:restriction},
we have $P(12)|_{UHF_{2}}=P[12]\oplus P[21]$ and
$P(1122)|_{UHF_{2}}=P[1122]\oplus P[2211]\oplus P[1221]\oplus P[2112]$.
Let $({\cal H},\pi)$ be $P(12)$ with the GP vector $\Omega$.
We denote $\pi(s_{i})$ by $s_{i}$ for simplicity of description.
By defining
$V_{1}\equiv \overline{\pi(UHF_{2})\Omega}$ and
$V_{2}\equiv \overline{\pi(UHF_{2})s_{2}\Omega}$,
we see that $(V_{1},\pi|_{UHF_{2}})$ and $(V_{2},\pi|_{UHF_{2}})$
are $P[12]$ and $P[21]$ with GP vectors $\Omega,s_{2}\Omega$ respectively.
Let $e_{1}\equiv \Omega$, $e_{2}\equiv s_{2}\Omega$ and
$t_{i}\equiv \psi_{12}(s_{i})$ for $i=1,2$.
Then we have
$t_{1122}e_{1}=e_{1},\quad t_{2112}e_{2}=e_{2},\quad
t_{1221}s_{1}e_{1}=s_{1}e_{1},\quad t_{2211}s_{2}e_{2}=s_{2}e_{2}$.
From $e_{1},s_{2}^{2}\Omega=s_{2}e_{2}\in V_{1}$ and $e_{2},s_{1}e_{1}\in V_{2}$,
we obtain $P[1122],P[2211]\prec V_{1}$ and $P[2112],P[1221]\prec V_{2}$.
Hence, by using $(P[12]\oplus P[21])\circ \psi_{12}=
P[1122]\oplus P[1221]\oplus P[2211]\oplus P[2112]$,
the statement for $\psi_{12}$ holds.

Next, by Table 2 and Theorem \ref{Thm:permurep} (iii), 
we have $P(12)\circ \psi_{13}=P(11)=P(1;+1)\oplus P(1;-1)$. 
From this result and Theorem \ref{Thm:restriction},
we obtain $(P[12]\oplus P[21])\circ \psi_{13}=P[1]\oplus P[1]$.
Therefore, the statement for $\psi_{13}$ holds.
\qedh
%
%
\begin{lem}
\label{lem:properties}
On $UHF_{2}$, the following holds.
\begin{enumerate}
\item
$\psi_{13},\psi_{12},\psi_{24}$ and $\psi_{34}$ are irreducible and proper.
\item
$\psi_{142}$ is reducible.
\end{enumerate}
\end{lem}
%
%
\pr
(i)
From $\alpha\circ \psi_{13}=\psi_{13}$,
$\psi_{13}(UHF_{2})$ is a subset of the fixed-point subalgebra
$(UHF_{2})^{\alpha}$ with respect to $\alpha$.
Hence, the image of $\psi_{13}$ is a proper subset of $UHF_{2}$.
On the other hand,  from (\ref{eqn:onethree}),
$\psi_{13}$ is irreducible.
Therefore the statement for $\psi_{13}$ holds.

Define the automorphism $\phi$ of $\co{2}$ by
$\phi(s_{1})\equiv (s_{1}+s_{2})/\sqrt{2}$,
$\phi(s_{2})\equiv (-s_{1}+s_{2})/\sqrt{2}$.
Then we have
%
%
\begin{equation}
\!\!\psi_{12}=({\rm Ad}(\phi\circ\alpha)) (\psi_{13}),\,\,
\psi_{24}=\psi_{13}\circ \alpha,\,\,
\psi_{34}=({\rm Ad}(\alpha\circ \phi\circ\alpha)) (\psi_{13}).
\end{equation}
Therefore, using that $\phi|_{UHF_{2}}\in{\rm Aut}UHF_{2}$,
the statements for $\psi_{12}$, $\psi_{24}$ and $\psi_{34}$ hold from
that for $\psi_{13}$.

\noindent (ii)
Let $\rho\equiv \psi_{142}$.
By the inductive method, we see that for any $x\in UHF_{2}$,
there exist
$y,z\in UHF_{2}$ such that $\rho(x)=s_{1}ys_{1}^{*}+s_{2}zs_{2}^{*}$.
For $a,b\in {\bf C}$, let $T_{a,b}\equiv aE_{11}+bE_{22}\in UHF_{2}$.
Then we see that $T_{a,b}\rho(x)=\rho(x)T_{a,b}$ for any $x\in UHF_{2}$.
Hence, $T_{a,b}\in \rho(UHF_{2})^{'}\cap UHF_{2}$ for each $a,b$.
Therefore the statement holds.
\qedh
Because the unitary equivalence in $E_{2,2}$ in Table 1 is given
by the unitary in $UHF_{2}$, there are at most
$16$ unitary equivalence classes in $UE_{2,2}$ 
in Theorem \ref{Thm:mainfour} with representatives in Table 2.
Furthermore, we can verify  the following identities in $UHF_{2}$:
%
%
\begin{equation}
\label{eqn:equations}
\psi_{14}=\psi_{1243},\quad \psi_{124}=\psi_{143},\quad
\psi_{132}=\psi_{234},\quad \psi_{23}=\psi_{1342}.
\end{equation}
Hence, there are at most $12$ unitary equivalence classes 
in $UE_{2,2}$ with representatives as follows:
%
%
\begin{equation}
\left\{\psi_{\sigma}:\sigma
=
\begin{array}{l}
id,\,(12)(34),\,(12),\,(13),\,(24),\, (34),\\
(142),\,(123),\,(14),\,(124),\,(132),\,(23)
\end{array}
\right\}.
\end{equation}
From Table 2, Proposition \ref{prop:uhf}, (\ref{eqn:sector}) and
Lemmas \ref{lem:branchingone}, \ref{lem:properties},
we obtain Table 3.
%
%
{\footnotesize
\[
\begin{array}{@{}cccccc@{}}
\multicolumn{6}{c}{\mbox{Table 3. Branching laws for $UE_{2,2}$.}}
\\
\toprule
\psi_{\sigma}& 
P[1]\circ\psi_{\sigma}
&
P[2]\circ\psi_{\sigma}
&
P[12]\circ\psi_{\sigma}
&
GP[+]\circ\psi_{\sigma}&
{\rm property}\\
\colrule
\psi_{id}      &P[1]&P[2]&P[12]& GP[+]&inn.aut\\
\psi_{(12)(34)}&P[2]&P[1]&P[21]& GP[+]&out.aut\\
\psi_{12}      &P[12]\oplus P[21]&P[1]\oplus P[2]&P[1122]\oplus P[2211]&\mbox{---}&irr.end\\\
\psi_{13}      &P[2]&P[2]&P[1]& \mbox{---}&irr.end\\
\psi_{24}      &P[1]&P[1]&P[2]& \mbox{---}&irr.end\\
\psi_{34}      &P[1]\oplus P[2]&P[12]\oplus P[21]&P[1221]\oplus P[2112]& \mbox{---}&irr.end\\
\psi_{142}     &P[12]\oplus P[21]&P[12]\oplus P[21]&P[1]\oplus P[2]& \mbox{---}
&red.end\\
\psi_{14}      &P[2]\oplus P[2]&P[1]\oplus P[1]&P[21]\oplus P[21]
&GP[+]\oplus GP[+]&red.end\\
\psi_{23}      &P[1]\oplus P[1]&P[2]\oplus P[2]&P[12]\oplus P[12]
&GP[+]\oplus GP[+]&red.end\\
\psi_{123}     &P[1]\oplus P[2]&P[1]\oplus P[2]&P[12]\oplus P[21]
&GP[+]\oplus GP[+]&red.end\\
\psi_{124}     &P[2]\oplus P[2]&P[1]\oplus P[1]&P[21]\oplus P[21]
&GP[+]\oplus GP[-]&red.end\\
\psi_{132}     &P[1]\oplus P[1]&P[2]\oplus P[2]&P[12]\oplus P[12]
&GP[+]\oplus GP[-]&red.end\\
\botrule
\end{array}
\]
}

\noindent
In order to show formulae of $GP[+]\circ \psi_{\sigma}$,
we use formulae $GP[\pm]\circ \theta=GP[\pm]$ 
$GP[+]\circ \beta_{1}=GP[+]\circ \beta_{2}=GP[-]$,
and Proposition \ref{prop:restriction} (ii), (\ref{eqn:sector}) and Table 2.
\\
\\
{\bf Proof of Theorem \ref{Thm:mainfour}.}
From (\ref{eqn:equations}) and $\#E_{2,2}=24$, we have $\#UE_{2,2}\leq 20$.
(ii) is verified by using the latter statements of (\ref{eqn:defeqn}) 
with a direct computation on $E_{JK}$'s.
Therefore, we obtain $\#(UE_{2,2}\setminus UG_{2})\leq 16$.
We classify them with respect to the image $\psi_{\sigma}(s_{1}s_{1}^{*})$ 
of $\psi_{\sigma}$ at $s_{1}s_{1}^{*}\in UHF_{2}$ in Table 4.
%
%
\vspace*{8pt}

\noindent
{\footnotesize
\begin{tabular}{cc|cc}
\multicolumn{4}{c}{Table 4. The image $\psi_{\sigma}(s_{1}s_{1}^{*})$ of $\psi_{\sigma}$ at $s_{1}s_{1}^{*}$.}
\\
\toprule
$\psi_{\sigma}(s_{1}s_{1}^{*})$& $\psi_{\sigma}$
&$\psi_{\sigma}(s_{1}s_{1}^{*})$& $\psi_{\sigma}$\\
\colrule
$s_{1}s_{1}^{*}$& $\psi_{12},\psi_{34}$&
$s_{1}s_{2}s_{2}^{*}s_{1}^{*}+s_{2}s_{1}s_{1}^{*}s_{2}^{*}$
& $\psi_{13},\psi_{123}$,\\
\cline{1-2}
$s_{2}s_{2}^{*}$&$\psi_{1324},\psi_{1423}$&&$\psi_{134}(\sim \psi_{142}),
\psi_{1234}(\sim \psi_{24})$\\
\colrule
$s_{1}s_{2}s_{2}^{*}s_{1}^{*}+s_{2}s_{2}s_{2}^{*}s_{2}^{*}$
& $\psi_{14},\psi_{124}$
&$s_{1}s_{1}s_{1}^{*}s_{1}^{*}+s_{2}s_{2}s_{2}^{*}s_{2}^{*}
$& $\psi_{24},\psi_{142}$,\\
\cline{1-2}
$s_{1}s_{1}s_{1}^{*}s_{1}^{*}+s_{2}s_{1}s_{1}^{*}s_{2}^{*}$& $\psi_{23},\psi_{132}$&&$\psi_{243}(\sim \psi_{123}),
\psi_{1432}(\sim \psi_{13})$\\
\botrule
\end{tabular}
}
\vspace*{8pt}

\noindent
Because, from Table 1 and Table 3,
endomorphisms in each case are not unitarily equivalent, 
they are different as elements in $UE_{2,2}$.
In consequence, we obtain that $\#(UE_{2,2}\setminus UG_{2})=16$.
This implies that $\#UE_{2,2}=20$ in (i). 
The latter in (i) follows from Table 3.
Likewise, (iii) also holds from Table 3.
\qedh

\noindent
We see that for any two endomorphisms in Table 3,
their branching laws are different.
\\

\vspace*{-5pt}
\noindent
{\it Notice}:
We summarize  remarkable results in this subsection.
(i) There is a proper irreducible endomorphism of $\co{2}$
such that its restriction on $UHF_{2}$ is not irreducible
(see $\psi_{142}$ in Table 1 and Table 3).
(ii) There is an endomorphism $\rho$ of $UHF_{2}$ such that
there are two different extensions of $\rho$ to $\co{2}$ 
and they are not equivalent in ${\rm End}\co{2}$
(compare Table 2 and (\ref{eqn:equations})).
(iii)
Table 3 is the complete classification
of endomorphisms in $UE_{2,2}$ with respect to
unitary equivalence.
Of course, this is properly finer than the classification
by statistical dimension \cite{Bau-Woll, DHR}.
Furthermore, we see that $4$ equivalence classes of 
irreducible and proper endomorphisms in $UE_{2,2}$.
In this sense, Table 3 contains nontrivial results.

%
%
\ssft{Nakanishi endomorphism restricted on $UHF_{3}$}
\label{subsection:thirdthree}
Our studies for endomorphisms were inspired by the following 
endomorphism $\rho_{\nu}$ of $\co{3}$ discovered by Noboru Nakanishi:
%
%
\begin{equation}
\label{eqn:nakeq}
\left\{
\begin{array}{c}
\rho_{\nu}(s_{1})\equiv 
s_{2}s_{3}s_{1}^{*}+s_{3}s_{1}s_{2}^{*}+s_{1}s_{2}s_{3}^{*},
\\ \\
\rho_{\nu}(s_{2})\equiv 
s_{3}s_{2}s_{1}^{*}+s_{1}s_{3}s_{2}^{*}+s_{2}s_{1}s_{3}^{*},
\\ \\
\rho_{\nu}(s_{3})\equiv 
s_{1}s_{1}s_{1}^{*}+s_{2}s_{2}s_{2}^{*}+s_{3}s_{3}s_{3}^{*},
\end{array}
\right.
\end{equation}
where $s_{1},s_{2},s_{3}$ are canonical generators of $\co{3}$.
In Theorem 1.2 of \cite{PE01}, 
we proved that $\rho_{\nu}$ is irreducible and
neither an automorphism nor equivalent to the canonical endomorphism of $\co{3}$.
By Table I (b) in \cite{PE02}, we obtained the following:
%
%
\begin{equation}
\label{eqn:nakanishi}
P(1)\circ \rho_{\nu}=P(3)\oplus P(12),\quad P(12)\circ \rho_{\nu}=P(113223).
\end{equation}

\noindent
Because $\rho_{\nu}$ is one of the second order permutative endomorphisms 
of $\co{3}$,
$\rho_{\nu}|_{UHF_{3}}$ is also an endomorphism of $UHF_{3}$.
%
%
\begin{prop}
%
%
\begin{eqnarray}
P[1]\circ \rho_{\nu}=&P[3]\oplus P[12]\oplus P[21],\\
P[12]\circ \rho_{\nu}=&P[113223]\oplus P[322311]\oplus P[231132],\\
P[21]\circ \rho_{\nu}=&P[223113]\oplus P[311322]\oplus P[132231].
\end{eqnarray}
\end{prop}
%
%
\pr
The first equation holds from (\ref{eqn:nakanishi})
and $P(12)|_{UHF_{3}}=P[12]\oplus P[21]$.
In the same way, we have
%
%
\begin{equation}
\begin{array}{rl}
(P[12]\oplus P[21])\circ \rho_{\nu}=&
P[113223]\oplus P[322311]\oplus P[231132]\\
&\oplus P[223113]\oplus P[311322]\oplus P[132231].
\end{array}
\label{eqn:branching}
\end{equation}

Let $({\cal H},\pi)$ be $P(12)$ of $\co{3}$ with the GP vector $\Omega$.
Then $e_{1}\equiv \Omega$ and $e_{2}\equiv s_{2}\Omega$ are 
GP vectors with respect to $P[12]$ and $P[21]$ 
in ${\cal H}$, respectively.
Define $V_{1}\equiv \overline{\pi(UHF_{3})e_{1}}$ 
and $V_{2}\equiv \overline{\pi(UHF_{3})e_{2}}$.
Then $e_{1},e_{2},s_{1}e_{1},s_{3}e_{1},s_{2}e_{2},s_{3}e_{2}$ are elements
of the cycle of $P(113223)$ in ${\cal H}$.
We see that $e_{1}\in V_{1},
s_{2}e_{2}=s_{2}s_{2}e_{1}=s_{22,21}e_{1}\in V_{1},
s_{3}e_{2}=s_{32,21}e_{1}\in V_{1}$,
$e_{2}\in V_{2},s_{1}e_{1}=s_{2}s_{1}e_{2}=s_{21,12}e_{2}\in V_{2},
s_{3}e_{1}=s_{31,12}e_{2}\in V_{2}$.
Let $t_{i}\equiv \rho_{\nu}(s_{i})$ for $i=1,2,3$.
Then we have
$t_{113223}e_{1}=e_{1}$, $t_{223113}e_{2}=e_{2}$,
$t_{311322}s_{1}e_{1}=s_{1}e_{1}$, $t_{132231}s_{3}e_{1}=s_{3}e_{1}$,
$t_{322311}s_{2}e_{2}=s_{2}e_{2}$ and $t_{231132}s_{3}e_{2}=s_{3}e_{2}$.
Therefore, we obtain
$P[113223],P[322311]$,
$P[231132]\prec V_{1}$ and $P[223113],P[311322]$, $P[132231]\prec V_{2}$.
From (\ref{eqn:branching}), the statement holds.
\qedh

%
%
\sftt{Branching laws on fermions}
\label{section:fourth}
%
%
%
\ssft{Fock representation and infinite wedge representation}
\label{subsection:fourthone}
The Fock representation and the infinite wedge representation
are well-known representations of fermions and
they are important not only in physics but also in mathematics \cite{MJD,Oko01}.
However there are few studies as a representation theory of 
the fermion algebra itself.
We review a relation between them and permutative representations \cite{IWF01}.
%
%
\begin{defi}
\label{defi:car}
\begin{enumerate}
\item
The Fock representation of ${\cal A}$ 
is the class of representation $({\cal H},\pi)$
with a cyclic vector $\Omega\in {\cal H}$ such that
%
%
\begin{equation}
\pi(a_{n})\Omega=0\quad(\mbox{for all } n\in {\bf N}).
\end{equation}
\item
The infinite wedge representation of ${\cal A}$ 
is the class of representation $({\cal H},\pi)$ 
with a cyclic vector $\Omega\in {\cal H}$ such that
%
%
\begin{equation}
b_{-k}\Omega= b_{k}^{*}\Omega=0 \quad(\mbox{for all }k\in{\bf Z}_{\geq}+1/2)
\end{equation}
where $b_{k}\equiv \pi(a_{2k+1}),\quad b_{-k}\equiv \pi(a_{2k})$
for  $k\in{\bf Z}_{\geq}+1/2$.
\item
$({\cal H}^{*},\pi^{*})$ is the dual of $({\cal H},\pi)$
if $({\cal H}^{*},\pi^{*})$ is equivalent to $({\cal H},\pi\circ \varphi)$ 
where $\varphi$ is the $*$-automorphism of ${\cal A}$ defined by
$\varphi(a_{n})\equiv (-1)^{n-1}a_{n}^{*}$ for each $n$.
\end{enumerate}
\end{defi}

\noindent
{\it Notice}: (a) 
The Fock representation and the infinite wedge representation are 
different from each other not only in numbering of generators 
but also in roles of creations and annihilations.
The infinite wedge representation is often called as the Fock representation
in a broad sense. In this paper, we distinguish them.
(b) The map $\varphi$ in Definition \ref{defi:car} (iii) is defined linearly
but not conjugate linearly. 
Hence, $({\cal H}^{*},\pi^{*})$ is not a conjugate representation
of $({\cal H},\pi)$.
The naming ``dual" is in conformity with
the dual infinite wedge in \cite{MJD}.

In Definition \ref{defi:car} (i) and (ii),
$\Omega$ in both cases is called the {\it vacuum vector}.
The creation and annihilation (operator) with respect to
their representations and vacua are given in Table 5.
%
%

{\footnotesize
\[
\begin{array}{@{}cccccccc@{}}
\multicolumn{8}{c}{\mbox{Table 5. Creations and anihilations.}}
\\
\toprule
&Fock&\quad &Fock^{*}&\quad &IW &\quad &IW^{*}\\
\colrule
\mbox{creation} &a_{n}^{*}&&a_{n}&& a_{2n-1}^{*},\,a_{2n}&&a_{2n-1},\,a_{2n}^{*}\\
\mbox{annihilation} &a_{n}&&a_{n}^{*}&&
a_{2n-1},\,a_{2n}^{*}&&a_{2n-1}^{*},\,a_{2n}\\
\botrule
\\
\multicolumn{8}{c}{
\begin{minipage}[c]{11cm}
Here $n\in {\bf N}$, and $Fock,Fock^{*},IW$ and $IW^{*}$ stand for
the Fock representation, the dual Fock representation,
the infinite wedge representation
and the dual infinite wedge representation
of ${\cal A}$, respectively.
\end{minipage}
}
\end{array}
\]
}

We regard ${\cal A}=UHF_{2}$ as the same subalgebra of $\co{2}$ by means of 
maps $\Phi_{CAR}$, $\Phi_{UHF_{2}}$, $\Psi$
in (\ref{eqn:recursive}), (\ref{eqn:mapone}), (\ref{eqn:psi}), respectively.
Restrictions of representations and branching laws
are also described by such identifications.

%
%
\begin{Thm}
\label{Thm:correspondence}
\hfill
\begin{enumerate}
\item
The (dual) Fock representation is $P[1]$ ({\it resp}. $P[2]$).
\item
The (dual) infinite wedge representation is $P[12]$ ({\it resp}. $P[21]$).
\item
In both (i) and (ii),
the vacuum vector of the former is the GP vector of the latter
up to scalar multiple.
\end{enumerate}
\end{Thm}
%
%
\pr
(i) 
In \cite{AK1}, we denoted $P(1)$ by ${\rm Rep}(1)$.
From $P(1)|_{UHF_{2}}=P(1)|_{CAR}=P[1]$ and $\S$ 3.3 of \cite{AK1}, 
the Fock representation is $P[1]$. 
From this result and $P[2]\circ \alpha=P[1]$, the statement holds.

\noindent (ii) This is shown in Proposition 3.6 (ii) of \cite{IWF01}.

\noindent (iii) 
Because of the uniqueness of the vacuum and GP vectors,
this is shown according to identification in (\ref{eqn:psi}).
\qedh

\noindent
By Theorem \ref{Thm:correspondence},
the four simplest examples $P[1],P[2],P[12],P[21]$ 
of permutative representations of $UHF_{2}$
are interpreted as well-known four representations in physics.
In this sense, the isomorphism $\Psi$, 
embeddings $\Phi_{CAR}$ and $\Phi_{UHF_{2}}$ are compatible with
permutative representations of $\co{2}$ and $UHF_{2}$, and
they acquire importance in mathematical physics.

%
%
\ssft{Permutative endomorphisms restricted on fermions}
\label{subsection:fourthtwo}
We show formulae of the second order permutative endomorphisms restricted 
on the CAR algebra ${\cal A}$.
We explain the computation method by using an example in $E_{2,2}$.

%
%
\begin{lem}
\label{lem:formulae}
For $n\geq 1$, the following holds:
%
%
\begin{eqnarray}
\psi_{142}(a_{2n-1})
=(-1)^{n-1}(a_{1}a_{1}^{*}a_{2n}-a_{1}^{*}a_{1}a_{2n}^{*}),\\
\psi_{142}(a_{2n})
=(-1)^{n-1}(a_{1}a_{1}^{*}a_{2n+1}^{*}+a_{1}^{*}a_{1}a_{2n+1}).
\end{eqnarray}
\end{lem}
%
%
\pr
Let $\rho\equiv \psi_{142}$.
First, we see that $\rho(a_{1})=s_{1}a_{1}s_{1}^{*}+s_{2}a_{1}^{*}s_{2}^{*}$ 
and $\rho(a_{2})=s_{1}a_{2}^{*}s_{1}^{*}-s_{2}a_{2}s_{2}^{*}$.
Next, we can verify that
if $X$ and $Y$ in $\co{2}$ satisfy that
$\rho(X)=s_{1}Xs_{1}^{*}+s_{2}X^{*}s_{2}^{*}$ and
$\rho(Y)=s_{1}Y^{*}s_{1}^{*}-s_{2}Ys_{2}^{*}$, 
then 
$\rho(\zeta(X))=s_{1}\zeta(X)^{*}s_{1}^{*}-s_{2}\zeta(X)s_{2}^{*}$ and
$\rho(\zeta(Y))=-s_{1}\zeta(Y)s_{1}^{*}-s_{2}\zeta(Y)^{*}s_{2}^{*}$
where $\zeta$ is defined in (\ref{eqn:rec}).
By the induction method, we obtain that
$\rho(a_{2n-1})=(-1)^{n-1}(s_{1}a_{2n-1}s_{1}+s_{2}a_{2n-1}^{*}s_{2}^{*})$ and
$\rho(a_{2n})=(-1)^{n-1}(s_{1}a_{2n}^{*}s_{1}-s_{2}a_{2n}s_{2}^{*})$.
Since we have $s_{1}Xs_{1}^{*}=a_{1}a_{1}^{*}\zeta(X)$ and
$s_{2}Xs_{2}^{*}=-a_{1}^{*}a_{1}\zeta(X)$ for any $X\in \co{2}$,
the statements hold.
\qedh

\noindent
In this way, we compute $\psi_{\sigma}(a_{n})$ for
every element in Table 3 and we show their properties in Table 6.

\noindent
%
%

{\footnotesize
\[
\begin{array}{ccc}
\multicolumn{3}{c}{\mbox{Table 6. Elements in $UE_{2,2}$ on fermions.}}
\\
\toprule
\psi_{\sigma}&\psi_{\sigma}(a_{n})&{\rm property}\\
\colrule
\psi_{id} &a_{n} &inn.aut\\
\psi_{(12)(34)}& a_{1}\quad (n=1),\quad (-1)^{n}a_{n}^{*}\quad(n\geq 2) &out.aut\\
\psi_{12} &\mbox{---} &irr.end\\
\psi_{13} &\mbox{---} &irr.end\\
\psi_{24} &\mbox{---} &irr.end\\
\psi_{34} &\mbox{---} &irr.end\\
\psi_{142}&(-1)^{k-1}(a_{1}a_{1}^{*}a_{2k}-a_{1}^{*}a_{1}a_{2k}^{*})\quad \quad
(n=2k-1)&red.end\\
&(-1)^{k-1}(a_{1}a_{1}^{*}a_{2k+1}^{*}+a_{1}^{*}a_{1}a_{2k+1})\quad(n=2k)&\\
\psi_{14} &(-1)^{n-1}(a_{1}a_{1}^{*}-a_{1}^{*}a_{1})a_{n+1}^{*} &red.end\\
\psi_{23} &(a_{1}a_{1}^{*}-a_{1}^{*}a_{1})a_{n+1} &red.end\\
\psi_{123}&a_{1}a_{1}^{*}a_{n+1}+(-1)^{n}a_{1}^{*}a_{1}a_{n+1}^{*} &red.end\\
\psi_{124}&(-1)^{n}(a_{1}^{*}-a_{1})a_{n+1}^{*} &red.end\\
\psi_{132}&(a_{1}^{*}-a_{1})a_{n+1} &red.end\\
\botrule
\end{array}
\]
}

\noindent
Because $\psi_{12},\psi_{13},\psi_{24}$ and $\psi_{34}$ 
are irreducible and proper,
they are important as nontrivial endomorphisms (or sectors) of ${\cal A}$.
We show their first three formulae in Table 7.

%
%
{\footnotesize
\[
\begin{array}{@{}cc@{}}
\multicolumn{2}{c}{\mbox{Table 7. $\psi_{\sigma}(a_1),\psi_{\sigma}(a_2), \psi_{\sigma}(a_3)$ for $\sigma=(12),(13),(24),(34)$.}}
\\
\toprule
\psi_{12}(a_1)=& - a_1 ( a_2 + a_2^* ) \\
\psi_{12}(a_2)=& -(a_1a_1^*a_2^{*} + a_1^*a_1a_2)(a_3 + a_3^*) \\
\psi_{12}(a_3) 
=&\{a_1a_1^*(a_2a_2^*a_3+a_2^*a_2a_3^*)
     -a_1^*a_1(a_2^*a_2a_3+a_2a_2^*a_3^*)\}(a_4+a_4^*)\\
\\
\psi_{13}(a_1)=&\,a_1^*a_2a_2^* + a_1a_2^*a_2\\
\psi_{13}(a_2)=&\,(a_1^*+a_1)(-a_2^*a_3a_3^* + a_2a_3^*a_3)\\
\psi_{13}(a_3)=&\,(a_1^*-a_1)(-a_2^*+a_2)(-a_3^*a_4a_4^* + a_3a_4^*a_4)\\
\\
\psi_{24}(a_1)=&\,a_1a_2a_2^* + a_1^{*}a_2^{*}a_2\\
\psi_{24}(a_2)
=&-(a_1^*+a_1)(-a_2a_3a_3^* + a_2^*a_3^*a_3)\\
\psi_{24}(a_3)
=&(a_1^*-a_1)(-a_2^*+a_2)(-a_3a_4a_4^* + a_3^{*}a_4^*a_4)\\
\\
\psi_{34}(a_1)=& - a_1 ( a_2 + a_2^* ) \\
\psi_{34}(a_2)=& -(a_1a_1^*a_2 + a_1^*a_1a_2^*)(a_3 + a_3^*) \\
\psi_{34}(a_3)=&\{-a_1a_1^*(a_2a_2^*a_3+a_2^*a_2a_3^*)
     +a_1^*a_1(a_2^*a_2a_3+a_2a_2^*a_3^*)\}(a_4+a_4^*)\\
\botrule
\end{array}
\]
}

\noindent
For Table 7, we use $\psi_{13}\circ \alpha=\psi_{24}$, 
$\alpha\circ \psi_{12}\circ \alpha=\psi_{34}$ and $\alpha(a_{n})=(-1)^{n-1}a_{n}^{*}$.
From Table 3 and Theorem \ref{Thm:correspondence}, we obtain branching laws 
on ${\cal A}$ in Table 8.

\noindent
%
%

{\footnotesize
\[
\begin{array}{@{}cccc@{}}
\multicolumn{4}{c}{\mbox{Table 8. Branching laws restricted for $UE_{2,2}$ on fermions.}}
\\
\toprule
\psi_{\sigma}& 
Fock\circ\psi_{\sigma}
&Fock^{*}\circ\psi_{\sigma}
&IW\circ\psi_{\sigma}\\
\colrule
\psi_{id}      &Fock&Fock^{*}&IW\\
\psi_{(12)(34)}&Fock^{*}&Fock&IW^{*}\\
\psi_{12}      &IW\oplus IW^{*}&Fock\oplus Fock^{*}&P[1122]\oplus P[2211]\\\
\psi_{13}      &Fock^{*}&Fock^{*}&Fock\\
\psi_{24}      &Fock&Fock&Fock^{*}\\
\psi_{34}      &Fock\oplus Fock^{*}&IW\oplus IW^{*}&P[1221]\oplus P[2112]\\
\psi_{142}     &IW\oplus IW^{*}&IW\oplus IW^{*}&Fock\oplus Fock^{*}\\
\psi_{14}      &Fock^{*}\oplus Fock^{*}&Fock\oplus Fock&IW^{*}\oplus IW^{*}\\
\psi_{23}      &Fock\oplus Fock&Fock^{*}\oplus Fock^{*}&IW\oplus IW\\
\psi_{123}     &Fock\oplus Fock^{*}&Fock\oplus Fock^{*}&IW\oplus IW^{*}\\
\psi_{124}     &Fock^{*}\oplus Fock^{*}&Fock\oplus Fock&IW^{*}\oplus IW^{*}\\
\psi_{132}     &Fock\oplus Fock&Fock^{*}\oplus Fock^{*}&IW\oplus IW\\
\botrule
\end{array}
\]
}
\vspace*{8pt}

\noindent
From Table 8, the Fock representation, the infinite wedge representation
and their duals are transformed to each other by endomorphisms. 
In case a branching occurs, new vacua appear instead of the original vacuum.
For example, we see that (\ref{eqn:composition}) is 
nothing but $\psi_{142}|_{{\cal A}}$. The branching law
$Fock\circ \psi_{142} =IW\oplus IW^{*}$ means 
the Fock vacuum is transformed to two vacua, that is,
the infinite wedge vacuum and the dual infinite wedge vacuum, by $\psi_{142}$.

%
%

\end{document}